\documentclass[12pt]{article}
\usepackage{oldgerm}
\usepackage{yfonts}
\usepackage{euler}
\usepackage{graphics,epstopdf}
\usepackage{bm}
\usepackage{graphicx}
\usepackage{amsmath}
\usepackage{amssymb}
\usepackage{amstext}
\usepackage{amscd}
\usepackage{amsfonts}
\usepackage{appendix}
\usepackage{color}
\DeclareMathAlphabet{\EuFrak}{U}{euf}{m}{n}
\DeclareMathAlphabet{\EuScript}{U}{eus}{m}{n}

\newcommand{\nd}{\noindent}

\newcommand{\be}{\begin{equation}}
\newcommand{\ee}{\end{equation}}
\newcommand{\ben}{\begin{eqnarray}}
\newcommand{\een}{\end{eqnarray}}

\title{{\bf A first order Tsallis theory}}

\author{{G. L. Ferri $^1$, A. Plastino$^2$, M. C. Rocca$^2$
and D. J. Zamora$^2$} \\
\small{$^1$Fac. de C. Exactas-National University La Pampa,} \\
\small{Peru y Uruguay, Santa Rosa, La Pampa, Argentina}\\
\small{$^2$ La Plata National University and
  Argentina's National Research Council}\\
\small{(IFLP-CCT-CONICET)-C. C. 727, 1900 La Plata - Argentina}}

\date{\today}

\begin{document}

\maketitle

\begin{abstract}

\nd We investigate first-order approximations to both i) Tsallis'
entropy $S_q$ and ii) the $S_q$-MaxEnt solution (called
q-exponential functions $e_q$). We use an approximation/expansion
for $q$ very close to unity. It is shown that the functions
arising from the procedure ii) are the MaxEnt solutions to the
entropy emerging from i). The present treatment is free of the
poles that, for classic quadratic Hamiltonians, appear in Tsallis'
approach, as demonstrated in [Europhysics Letters {\bf 104},
(2013), 60003]. Additionally, we show that our treatment is
compatible with extant date on the ozone layer.

\end{abstract}

\nd Keywords: MaxEnt, second variation, generalized statistics.
Tsallis entropy, divergences,  partition function,
specific heat.

\newpage

\renewcommand{\theequation}{\arabic{section}.\arabic{equation}}

\section{Introduction}

\setcounter{equation}{0}

 During the last quarter of century, an active subfield of
statistical mechanics is centered  around the concept of the
so-called q-statistics, that  Tsallis introduced in
\cite{tsallis88,tsallis}, that appears to yield better answers, in
many scenarios, than the orthodox Boltzmann-Gibbs  entropic
functional \cite{tsallisbook}. These scenarios involve variegated
disciplines (see, for instance, \cite{[4]}-\cite{[44]},  \cite{[5]},
\cite{[6]},
 \cite{[7]},  \cite{[8]}, \cite{[9]}, \cite{[10]},  \cite{[11]}, \cite{[12]}, \cite{[13]},
 \cite{[14]}, \cite{wilk}, etc.) Concepts involving q-statistics are important not only in physics
but  in chemistry, biology, mathematics, economics, and informatics
as well \cite{[18]}, \cite{[19]}.

\vskip 3mm
  In this work we revisit the Tsallis-subject by appealing to perturbation theory around
  $q=1$. We investigate first-order approximations to both A) Tsallis'
entropy $S_q$ and B) the $S_q$-MaxEnt solution (called q-exponential
functions $e_q$). It is shown that the functions arising from the
procedure B) are the MaxEnt solutions to the entropy emerging from
A). The present treatment is free of the poles that, for classic
quadratic Hamiltonians, plague Tsallis' approach, as demonstrated in
\cite{epl}. Additionally, we show that our treatment is compatible
with extant date on the ozone layer \cite{nasa,Tsallis04,gustavo}.

\subsection{Motivation}

It was shown in  \cite{vp} that  data-detection following
 a  normalization step does not permit straightforward  inference of
 data-distribution in exponential or Gaussian  fashion because
 of a  systematic transformation into $q-$exponentials or
 q-Gaussians.   The origin of the often encountered $q-$exponential or q-Gaussian  data needs
careful analysis. For a  very large set of recorded-data (elliptical
ones), this occurrence is a simple consequence of a
device-normalization stage \cite{vp}. This entails that the
q-neighborhood  of $q=1$ is extremely important for q-statistics,
deserving the special attention that it receives below.

\nd Note also that  in the superstatistics approach  of Beck and Cohen \cite{cohen}, the parameter
$q-1$ is a measure of temperature fluctuations in the driven nonequilibrium system with a stationary state, so small $q-1$ corresponds to sharply peaked temperature distributions around the mean.

\setcounter{equation}{0}

\section{Still another new entropy}

After the pioneer Tsallis' paper \cite{tsallis}, many new entropies
have been proposed \cite{landsberg}. In this vein, we begin our
considerations with reference to a new q-entropy, that exhibits over
Tsallis' one some important advantages to be discussed below. The
new entropy will emerge as a result of a the first order
approximation (around $q=1$) of the q-exponential function
\cite{tsallisbook}:
\begin{equation}
\label{er2.1} [1+(1-q)\beta U]^{\frac {1} {q-1}}\simeq \left[1+\frac
{(1-q)} {2}\beta^2U^2\right] e^{-\beta U}.
\end{equation}
The pertinent probability distribution becomes now, instead of the
q-exponential \cite{tsallisbook}, the following one

\begin{equation}
\label{er2.2} P= \frac {\left[1+\frac {(1-q)} {2}\beta^2U^2\right]
e^{-\beta U}} {{\cal Z}},
\end{equation}
with
\begin{equation}
\label{er2.3} {\cal Z}=\int\limits_{M} \left[1+\frac {(1-q)}
{2}\beta^2U^2\right] e^{-\beta U} d\mu.
\end{equation}
We construct next the first order approximation to Tsallis' entropy
\cite{tsallisbook} and find
\begin{equation}
\label{er2.4} {\cal S}_q=\frac {1} {1-q}\left(1-\int\limits_M
P^qd\mu\right)\simeq -\int\limits_ MP\ln P\left[ 1+\frac {(q-1)}
{2}\ln P\right]d\mu.
\end{equation}
We show next (\ref{er2.2}) arises from extremizing (\ref{er2.4}).
 The ensuing variational problem revolves around a Lagrangian

\[{\cal F}_{{\cal S}_q}(P)=
-\int\limits_ MP\ln P\left[
1+\frac {(q-1)} {2}\ln P\right]d\mu+\lambda_1\left(
\int\limits_MPUd\mu-<U>\right)+\]
\begin{equation}
\label{er2.5} \lambda_2\left( \int\limits_MPd\mu-1\right),
\end{equation}
whose increment is
\[{\cal F}_{{\cal S}_q}(P+h)=
-\int\limits_ M(P+h)\ln(P+h)\left[
1+\frac {(q-1)} {2}\ln(P+h)\right]d\mu+\]
\begin{equation}
\label{er2.6} \lambda_1\left( \int\limits_M(P+h)Ud\mu-<U>\right)+
\lambda_2\left( \int\limits_M(P+h)d\mu-1\right),
\end{equation}
so that
\[{\cal F}_{{\cal S}_q}(P+h)-{\cal F}_{{\cal S}_q}(P)=
-\int\limits_ M(P+h)\ln(P+h)\left[
1+\frac {(q-1)} {2}\ln(P+h)\right]d\mu+\]
\begin{equation}
\label{er2.7} \int\limits_ MP\ln P\left[ 1+\frac {(q-1)} {2}\ln
P\right]d\mu+ \lambda_1 \int\limits_MUhd\mu+ \lambda_2
\int\limits_Mhd\mu,
\end{equation}
or
\[{\cal F}_{{\cal S}_q}(P+h)-{\cal F}_{{\cal S}_q}(P)=
-\int\limits_ M\left[1+\ln P+
\left(\frac {q-1} {2}\right)
\left(2\ln P+\ln^2P\right)-\lambda_1
-\lambda_2\right]hd\mu\]
\begin{equation}
\label{er2.8} -\int\limits_ M\left[\frac {1} {2P}+ \left(\frac {q-1}
{2}\right) \left(\frac {1+\ln P} {P}\right)\right]h^2d\mu.
\end{equation}
The last relation yields the extremizing distribution $P$
\begin{equation}
\label{er2.9}
1+\ln P+
\left(\frac {q-1} {2}\right)
\left(2\ln P+\ln^2P\right)-\lambda_1
-\lambda_2=0
\end{equation}
\begin{equation}
\label{er2.10} -\int\limits_ M\left[\frac {1} {P}+ \left(q-1\right)
\left(\frac {1+\ln P} {P}\right)\right]h^2d\mu\leq C||h||^2,
\end{equation}
with  $C<0$ is a  constant.
See \cite{shilov}, \cite{max}.
Replacing  (\ref{er2.2}) into
(\ref{er2.9}) one verifies that $P$ is a solution to (\ref{er2.9}),
with  $\lambda_1$ and  $\lambda_2$ given by
\begin{equation}
\label{er2.11}
\lambda_1=-\beta[q-(q-1)\ln{\cal Z}]
\end{equation}
\begin{equation}
\label{er2.12}
\lambda_2=1-q\ln{\cal Z}+\left(\frac {q-1} {2}\right)\ln^2{\cal Z}
\end{equation}
 Of course,  (\ref{er2.10}) must be verified for a maximum. \vskip
 3mm

We realize that the entropy  (\ref{er2.4}) is not just an
approximation but a legitimate new thermodynamic one, since it
complies with the MaxEnt strictures.

\subsection{Comparison between the exact and approximate solutions}

Figs. 1,2,3, and  4 correspond to the modulus of the ratio between the approximate and the exact solutions (AER), eq. (\ref{er2.1}).  
Horizontal coordinates  are in meters. 
For simplicity we have taken $U=x^2$ and $\beta=1$.
The agreement is excellent. More to the point,
it is excellent over extremely long distances,
for atomic phenomena.

\begin{figure}
\includegraphics[width=12cm]{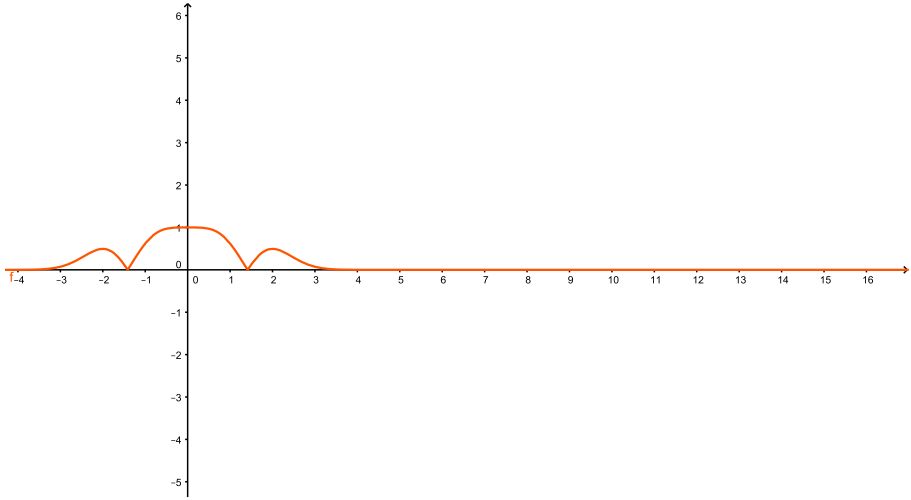}
\caption{AER (see text) for  $1-q=0.5$} \label{tres}
\end{figure}

\begin{figure}
\includegraphics[width=12cm]{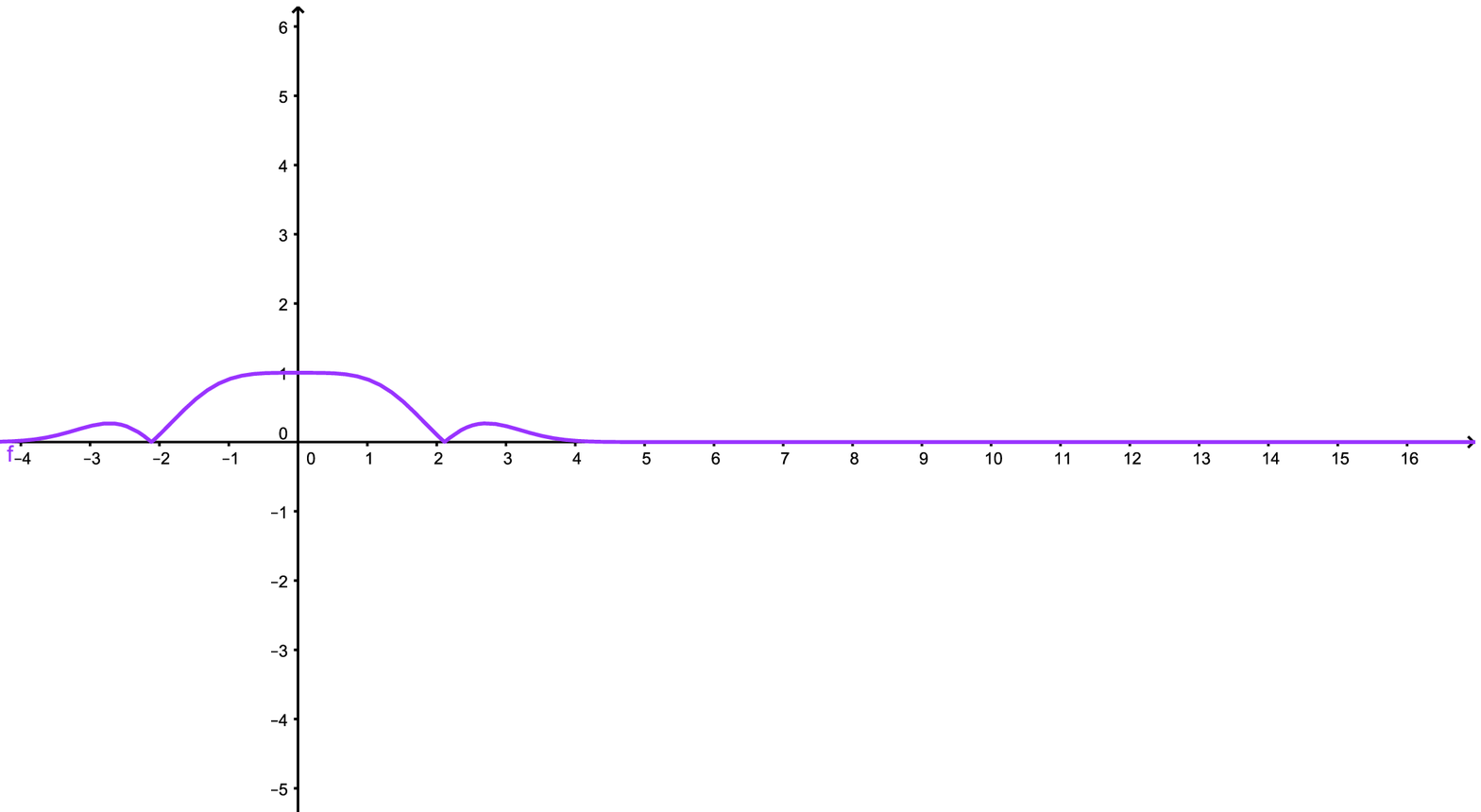}
\caption{AER (see text) for $1-q=0.1$} \label{cuatro}
\end{figure}

\begin{figure}
\includegraphics[width=12cm]{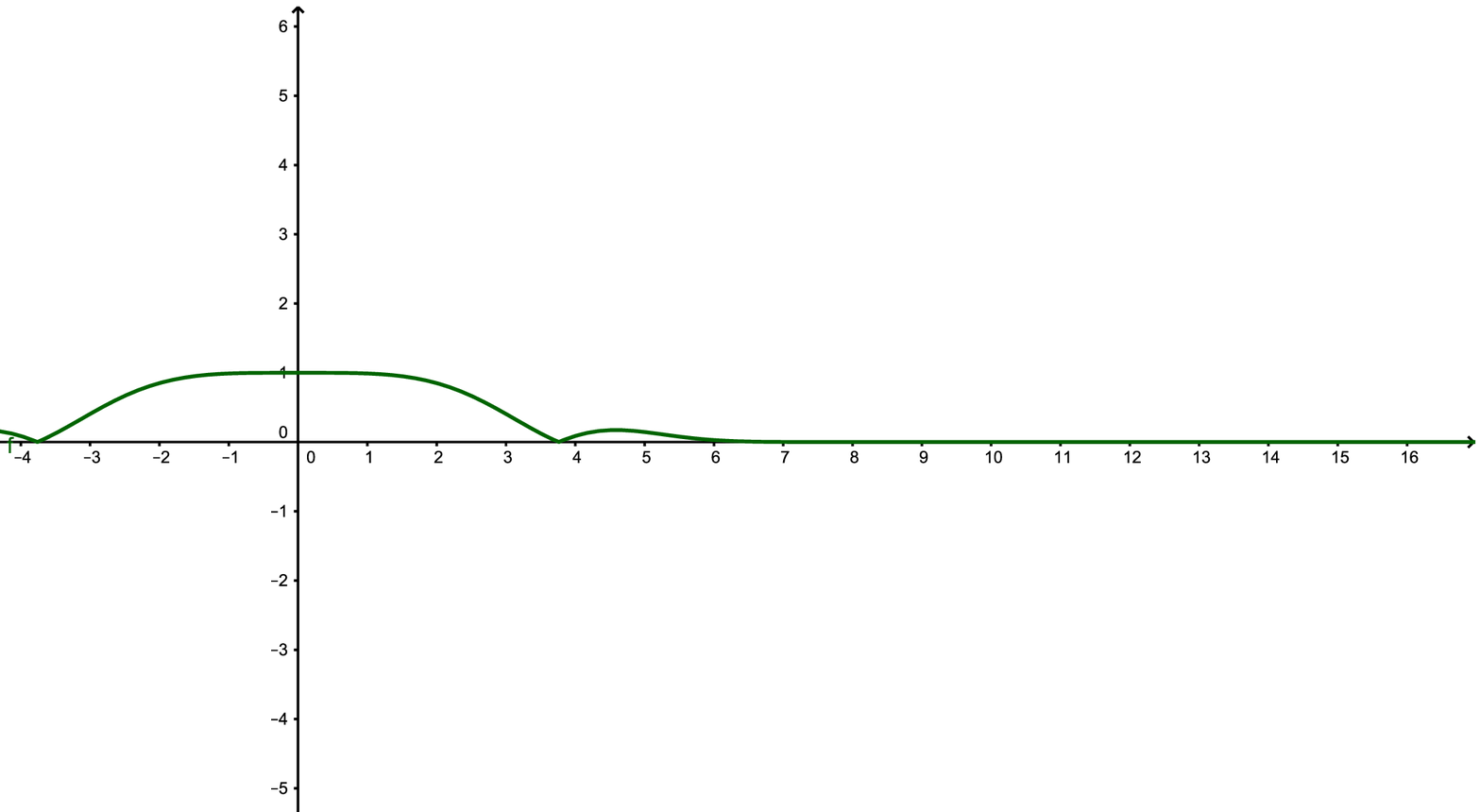}
\caption{AER (see text) for $1-q=0.01$} \label{cinco}
\end{figure}

\begin{figure}
\includegraphics[width=12cm]{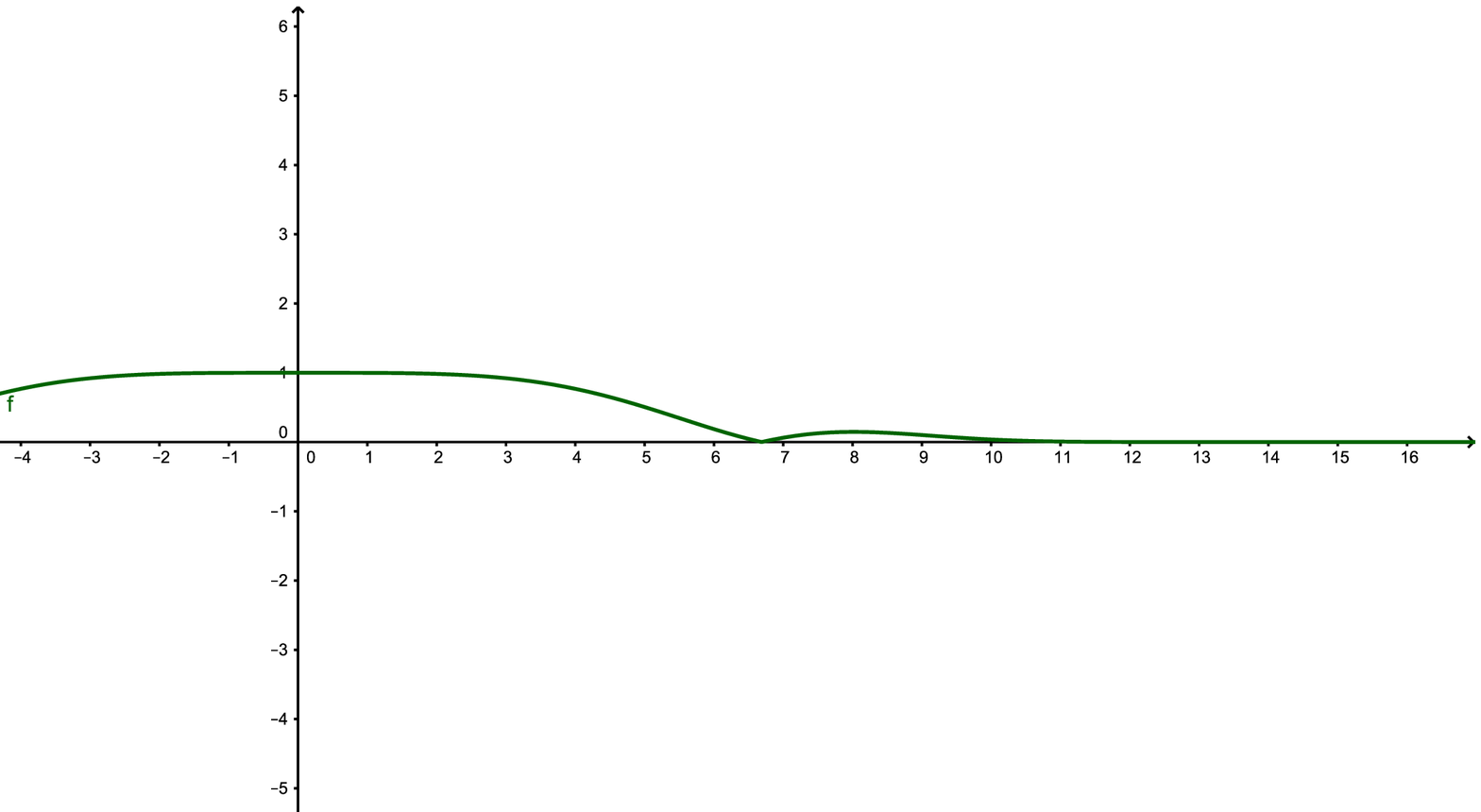}
\caption{AER (see text) for $1-q=0.001$ } \label{seis}
\end{figure}

\newpage

\section{Quadratic  Hamiltonians}
\subsection{Review of Tsallis' treatment}
\setcounter{equation}{0}

In  reference \cite{epl} one finds the associated  partition
function and mean energy for Tsallis'   q-MaxEnt approach, i.e.,

\begin{equation} \label{ep2.1} {\cal
Z}=\frac {\pi^{\nu}} {\Gamma(\nu)} \int\limits_0^{\infty} \frac
{u^{\nu-1}} {[1+\beta(q-1)u]^{\frac {1} {q-1}}} du,
\end{equation}
where the  integral is evaluated using  \cite{grad}:
\begin{equation}
\label{ep2.2} {\cal Z}= \frac {\pi^{\nu}} {[\beta(q-1)]^{\nu}} \frac
{\Gamma\left(\frac {1} {q-1}-\nu\right)} {\Gamma\left(\frac {1}
{q-1}\right)}.
\end{equation}
This result is valid for $q\neq 1$ and we have selected $1\leq q
<2$. Of course,  $q=1$ is the  orthodox result, for which the
q-exponential transforms itself into the ordinary exponential
function (and the integral (\ref{ep2.1}) is convergent). The
singularities (divergences) of (\ref{ep2.1}) are, of course,  given
by the poles of the $\Gamma$ function, that is,  for
\[\frac {1} {q-1}-\nu=-p\;\;{\rm for} \;\;p=0,1,2,3,......,\]
  i.e., $q-$ values given by
\[q=\frac {3} {2},\frac {4} {3},\frac {5} {4},\frac {6} {5},......,
\frac {\nu} {\nu-1},\frac {\nu+1} {\nu}\] For the mean enrgy,
instead,  (\ref{ep2.1}) gives
\begin{equation}
\label{ep2.3} <{\cal U}>=\frac {\pi^{\nu}} {\Gamma(\nu){\cal Z}}
\int\limits_0^{\infty} \frac {u^{\nu}} {[1+\beta(q-1)u]^{\frac {1}
{q-1}}} du,
\end{equation}
so that,  using \cite{grad} one finds
\begin{equation}
\label{ep2.4} <{\cal U}>=\frac {\nu\pi^{\nu}} {{\cal
Z}[\beta(q-1)]^{\nu+1}} \frac {\Gamma\left(\frac {1}
{q-1}-\nu-1\right)} {\Gamma\left(\frac {1} {q-1}\right)},
\end{equation}
Here, the poles  are given by
\[\frac {1} {q-1}-\nu-1=-p\;\;\;{\rm for}\;p=0,1,2,3,......,\]
or,
\[q=\frac {3} {2},\frac {4} {3},\frac {5} {4},\frac {6} {5},......,
\frac {\nu+1} {\nu},\frac {\nu+2} {\nu+1}.\] As customary
\cite{2PP93}, using q-logarithms \cite{tsallis} $\ln_q(x)=
\frac{x^{1-q} - 1 }{1-q}$, Tsallis'  entropy becomes
\begin{equation}
\label{ep2.5} {\cal S}_q=\ln_q {\cal Z} + {\cal Z}^{1-q} \beta
<{\cal U}>,
\end{equation}
that is finite if ${\cal Z}$ and $<{\cal U}>$ are also finite.

\subsection{The new entropy alternative}

\setcounter{equation}{0}

The new partition function is easily seen to be
\begin{equation}
\label{ep3.1} {\cal Z}=\frac {\pi^{\nu}} {\Gamma(\nu)}
\int\limits_0^{\infty} u^{\nu-1} \left[1+\frac {(q-1)}
{2}\beta^2u^2\right] e^{-\beta u}du,
\end{equation}
and, evaluating the integral,

\begin{equation}
\label{ep3.2} {\cal Z}=\frac {\pi^{\nu}} {\beta^{\nu}} \left[1+\frac
{(q-1)\nu(\nu+1)} {2}\right].
\end{equation}
No poles are detected! For  $q=1$ this yields the  Boltzmann-Gibbs'
(BG) partition function. For the mean energy one has

\begin{equation}
\label{ep3.3} <{\cal U}>=\frac {\pi^{\nu}} {\Gamma(\nu){\cal Z}}
\int\limits_0^{\infty} u^{\nu} \left[1+\frac {(q-1)}
{2}\beta^2u^2\right] e^{-\beta u}du,
\end{equation}
and after  integration
\begin{equation}
\label{ep3.4} <{\cal U}>=\frac {\nu\pi^{\nu}} {\beta^{\nu+1}{\cal
Z}} \left[1+\frac {(q-1)} {2}(\nu+1)(\nu+2)\right],
\end{equation}
with, again, no poles. Using now  (\ref{ep3.2}) we obtain

\begin{equation}
\label{ep3.5} <{\cal U}>=\frac {\nu} {\beta}
\left[1+(q-1)(\nu+1)\right],\end{equation} that coincides with the
BG result for   $q=1$. As for the entropy, one must develop up to
first order ${\cal Z}^{1-q}$ and we get

\begin{equation}
\label{ep3.6} {\cal Z}^{1-q}=1+(q-1)\nu\ln\left(\frac {\beta}
{\pi}\right)+ \frac {(q-1)^2} {2}\left[\nu^2\ln^2\left(\frac {\beta}
{\pi}\right)- \nu(\nu+1)\right],
\end{equation}
that together with  (\ref{ep2.5}) leads to

\begin{equation}
\label{ep3.7} {\cal S}_q=\nu\left[1+\ln\left(\frac {\beta}
{\pi}\right)\right]+ (q-1)\left[\nu+1-\frac {\nu(\nu+1)} {2}+
\nu\ln\left(\frac {\beta} {\pi}\right)- \frac {\nu^2}
{2}\ln^2\left(\frac {\beta} {\pi}\right)\right],
\end{equation}
that for $q=1$ is the BG result.

\subsection{Specific Heat}

\setcounter{equation}{0}

We need the derivative of (\ref{ep3.5}) with respect to the
temperature $T$ to reach
\begin{equation}
\label{ep4.1} {\cal C}=\nu{\cal K} \left[1+(q-1)(\nu+1)\right],
\end{equation}
with  ${\cal K}$  Boltzmann's constant. For $q=1$ we reobtain the BG
result. The corrections in \ref{ep4.1} to the BG could easily be
checked out empirically.

\section{The Ozone layer}

\setcounter{equation}{0} Tsallis' $q-$triplet \cite{Tsallis04} is
possibly the most spectacular empirical quantifier of
non-extensivity, i.e., $q \ne 1$. The quantifier was  studied in
\cite{gustavo}  with reference to an experimental time-series
related to the daily depth-values of the stratospheric ozone layer.
Pertinent data were there  expressed in Dobson units and ranged from
1978 till 2005. After evaluation of the three associated Tsallis'
q-indexes one concluded that nonextensivity is clearly a
characteristic of the ozone layer.

\vskip 3mm

Stratospheric ozone is encountered mainly within a $\sim 15$km-layer
at a height of about 15km. There is a low density of a few
$O_3-$molecules per million of air-molecules.  The associated
mechanism of  interactions responsible for depletion is given  in
Ref. \cite{nasa}.  A stationary regime prevails,  modulated by
various types  of oscillations, that is 1) a yearly one
  due to the orientation of the incoming radiation,
 2) other  of a period of around 2 years originated in  stratospheric air-currents, and 3)
 a secular variation \cite{nasa}. In \cite{gustavo} the authors concentrated efforts on
 two time-series: A) $\{Z_n\}$ of depth-values for the ozone
 layer and B) its daily variability $\{\Delta Z_n\}$.

 \vskip 3mm

  Tsallis' theory  displays three important q-features (three different q-values)   \cite{Tsallis04}:
\begin{itemize}
\item i)  A q-value linked to
meta-stable  states, the one of the pertinent $q$-exponential, that
we call  $q\equiv q_{stat}$.

\item ii) The above states display a $q$-exponential sensibility to
initial conditions (the so-called weak chaos). We speak of a q-value
that we call  $q_{sens}$.

\item  iii) Meta-stable macroscopic quantities relax to their $q=1$-values  in
a
 $q$-exponential fashion with $q=q_{rel}$. \end{itemize}

Thus, a meta-stable state is characterized by a triplet of
$q-$values:  $(q-{stat}, q-{sens}, q-{rel})\neq(1, 1, 1)$, where
$q-{stat}>1$, $q-{sen }<1$, and $q-{rel}>1$ \cite{Tsallis04}.

 \vskip 3mm

Since in the case of the BG statistics the three different q-values
above coalesce to $q=1$, with the present treatment we expect a
convergence of the three triplet's q-values to just one value close
to unity. Our numerical results, computed here following the
methodology described in \cite{gustavo}, do not falsify this
convergence. This is a rather important numerical result. We
evaluated $q-stat$ and $q-rel$ for our comparison.  ($q-{sen }$
implies a much more involved calculation.) Note that the q-values
are determined by the ozone-data. We use satellite-data
corresponding to Buenos Aires city. These are daily values $z_n$
obtained from November '78 till May '93 and from July '96 till Dec.
'05. \vskip 3mm

To calculate $q_{stat}$ we adjust the histogram with a $q$
-Gaussian. The one that fits best data is a q-Gaussian q = 1.32. In
the case of our  first order treatment, we use a "fist order
q-Gaussian" \be p(z)=[1+\frac{(1-q)}{2}a^2 z^4]e^{-az^2}, \ee
properly normalized, of course.

The correlation curve has been adjusted with a $q$-Gaussian with
q=1.888 and in the first order case we use \be
p(z)=[1+\frac{(1-q)}{2}\beta^2 z^2]e^{-\beta z }, \ee again properly
normalized.

The suitable $q-$value for the stationary state is obtained from the
probability distribution function (PDF) [here either Gaussian-$e_q$
or the MaxEnt PDF \ref{er2.2}], associated to daily variations of
the ozone layer's depth $\Delta Z_n = Z_{n+1}-Z_n$. This $\Delta
Z-$range is subdivided into little cells of width (in Dobson units
(UD)) $\delta z$, centered at $z_i$, so that one can assess with
which frequency $\Delta Z-$values fall within each cell. We chose a
cell-size $\delta z=5 UD$. The resultant histogram, properly
normalized, gives our stationary-PDF $\{p\,(z_i)\}_{i=1}^N$. Of
course, $p_i$ is the probability for a
 $\Delta Z-$value to fall within the  $i$th cell, centered at $z_i$,
 with $N$ the cell-number  \cite{gustavo}.
We have

\begin{enumerate}

\item Tsallis' difference: $|q-rel - q-stat|= 0.57$

\item Our difference: $|q-rel - q-stat|= 0.08$, much smaller than
the preceding one.

\end{enumerate}

Figure 1 illustrates the statistical q-situation. Red circles yield
the histogram data. The black curve displays the best fit to the
data  for a q-Gaussian
 and the blue one our MaxEnt PDF \ref{er2.2}.

\begin{figure}
\includegraphics[width=12cm]{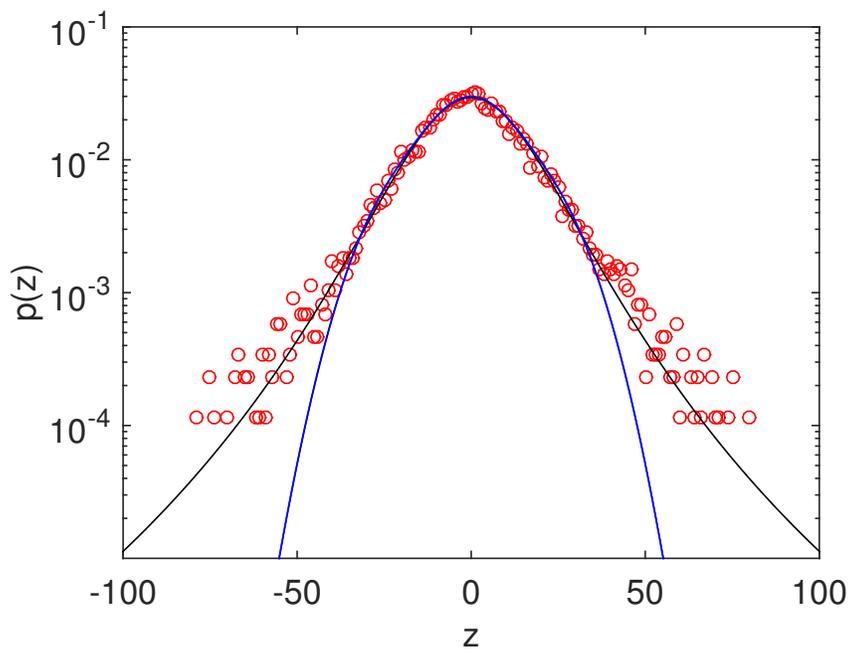}
\caption{Red circles correspond to histogram data $p(z)$ vs. $z$;
solid black line: the $q$-Gaussian
function that fits $p(z_i)$; Blue curve:
the best adjustment with the first order q-Gaussian
properly normalized} \label{uno}
\end{figure}

The $q-rel-$value is determined via the temporal
self-correlation coefficient

\be C(\tau)=\frac{\sum_n Z_{n+\tau}.Z_n}{\sum_n Z_n^2}.\ee  For a
classical BG-process such correlation should decay in exponential
fashion, which is not the case for our data. Fig. 2 refers to q-rel.
Black circles correspond to the correlation for distinct $\tau$.
Black curve: best q-Gaussian-fit to the data and red curve, same for
our PDF (1.22).

\begin{figure}
\includegraphics[width=12cm]{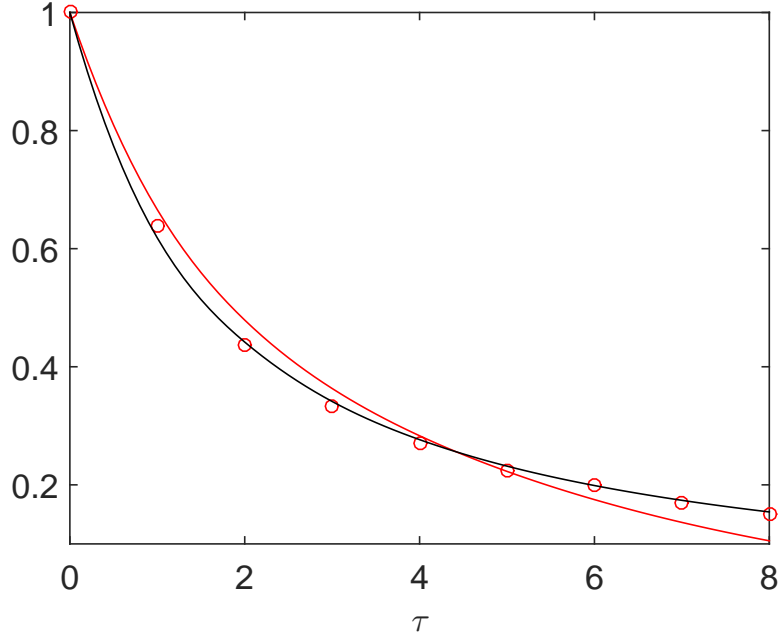}
\caption{$\ln_q$ of the self correlation coefficient $C(\tau)$ vs.
time delay $\tau$ (in days). The linear CC is 0.999. Black curve is
for a q-Gaussian. Red curve, same for our PDF
(1.22).}\label{dos}\end{figure} \newpage

\section{Conclusions}

In this effort we have investigated first-order approximations to
both 1) Tsallis' entropy $S_q$ and 2) the $S_q$-MaxEnt solution
(called q-exponential functions $e_q$). We were able to  show that
the functions arising from the MaxEnt treatment 2)  are precisely
the MaxEnt solutions to the approximate entropy arising  from 1).
This entails that the approximate entropy is a legitimate new
entropic functional.

The present treatment with he new entropy is free of the poles that,
for classic quadratic Hamiltonians, appear in Tsallis' approach, as
demonstrated in [Europhysics Letters {\bf 104}, (2013), 60003], in
both the partition function and the mean energy.

We showed that our treatment is compatible with extant date
on the ozone layer. The associated q-triplet \cite{Tsallis04}
Tsallis' $q-$triplet \cite{Tsallis04} is perhaps the most
spectacular empirical quantifier of non-extensivity, i.e., $q \ne
1$. The quantifier was  studied in \cite{gustavo} for Tsallis'
entropy, and we see that the present new q-entropy can accommodate
the triplet phenomenon.

Finally, we emphasize that the main idea of the current paper is based 
on the approximation of Eq. (2.1). There is in leading order a quadratic correction term in
the variable $U$  if the vicinity of the ordinary Boltzmann factor is considered in the q-statistics approach. This quadradic correction term  has been discussed also in \cite{cohen}, where it was also found  that the results for small $q-1$ are universal, i.e. applicable to many physical situations in the same way. What is actually new in the current effort is  to
promote these small $q-1$ effects to yield a new MaxEnt formalism.

\vskip 6mm

{\bf Acknowledgment:} We thank support from Conicet's PIP 029/12.


\newpage

\begin{thebibliography}{99}


\bibitem{tsallis88} C. Tsallis, J. of Stat. Phys., {\bf 52}
(1988) 479.


\bibitem{tsallis} C. M. Gell-Mann and C. Tsallis, Nonextensive
Entropy�Interdisciplinary Applications (Oxford University Press, New
York, 2004).

\bibitem{tsallisbook} C. Tsallis, {\it Introduction to Nonextensive Statistical Mechanics � Approaching a Complex World}
 (Springer, NY, 2009).



\bibitem{[4]} A. Adare et al., Phys. Rev. D {\bf 83} (2011) 052004.

\bibitem{[44]} G. Wilk, Z. Wlodarczyk, Physica A {\bf 305} (2002) 227.



\bibitem{[5]} R. M. Pickup, R. Cywinski, C. Pappas, B. Farago, and P.
Fouquet, Phys. Rev. Lett. 102 (2009)   097202.

\bibitem{[6]} E. Lutz and F. Renzoni, Nature Physics {\bf 9}   (2013)    615.

\bibitem{[7]} R. G. DeVoe, Phys. Rev. Lett. {\bf 102}  (2009)     063001.

\bibitem{[8]} Z. Huang, G. Su, A. El Kaabouchi, Q. A. Wang, and J. Chen, J.
Stat. Mech.  L05001 (2010).

\bibitem{[9]} J. Prehl, C. Essex, and K. H. Hoffman, Entropy {\bf 14}  (2012)     701.


\bibitem{[10]} B. Liu and J. Goree, Phys. Rev. Lett. {\bf 100}   (2018)    055003.


\bibitem{[11]} O. Afsar and U. Tirnakli, EPL {\bf 101}   (2013)    20003.

\bibitem{[12]} U. Tirnakli, C. Tsallis, and C. Beck, Phys. Rev. E {\bf 79}  (2009)
056209.

\bibitem{[13]} G. Ruiz, T. Bountis, and C. Tsallis, Int. J. Bifurcation
Chaos {\bf 22}   (2012)    1250208.

\bibitem{[14]} C. Beck and S. Miah, Phys. Rev. E {\bf 87}  (2013)    031002.

\bibitem{[15]} A. A. Budini, Phys. Rev. E {\bf 86}  (2012)     011109.

\bibitem{[16]} J.-L. Du, J. Stat. Mech.  P02006 (2012).

\bibitem{wilk}
G. Wilk, Z. Wlodarczyk, Phys. Rev. Lett. {\bf 84}    (2000) 2770.



\bibitem{[18]} S. Abe, Astrophys. Space Sci. 305, 241 (2006).

\bibitem{[19]} S. Picoli, R. S. Mendes, L. C. Malacarne, and R. P. B.
Santos, Braz. J. Phys. 39, 468 (2009).




\bibitem{epl} A. Plastino and M. C. Rocca
Europhysics Letters {\bf 104}, (2013), 60003.

\bibitem{shilov} G. Y. Shilov: {\it Mathematical Analysis}
(Pergamon Press, NY, 1965).

\bibitem{max} A. Plastino and M. C. Rocca
Physica A {\bf 436}, (2015), 572.

\bibitem{nasa} R.M. Todaro Ed., \emph{Stratospheric ozone}, NASA's Goddard
Space Flight Center Atmospheric Chemistry and Dynamics Branch.
http://www.ccpo.odu.edu/ SEES/ozone/oz\_class.htm


\bibitem{Tsallis04} C. Tsallis,
Physica A \textbf{340} (2004) 1.

\bibitem{gustavo} G.L. Ferri, M.F. Reynoso Savio, A. Plastino, Physica A {\bf 389} (2010) 1829.


\bibitem{vp} C. Vignat, A. Plastino.
Physica A {\bf 388} (2009) 601.

\bibitem{cohen} C. Beck, E.G.D. Cohen, Physica A {\bf 322} (2003)  267.


 \bibitem{landsberg} P. T. Landsberg, Braz. J. Phys. {\bf 29} (1999) 46.


\bibitem{grad} I. S. Gradshteyn and I. M. Rizhik, {\it Table
of Integrals Series and Products} (Academic Press, NY, 1965,
p.285, {\bf 3.194},3).

\bibitem{2PP93} A. R.  Plastino, A Plastino
Physics Letters A {\bf 177} (1993) 177.

\bibitem{xxx} The procedure we have employed here  is usually called
``dimensional regularization".




\bibitem{PP94} A. R. Plastino, A. Plastino, Phys. Lett. A  {\bf 193}  (1994) 140.





\bibitem{witten} J. J. Atick, E.  Witten, Nucl. Phys. B {\bf 310} (1988) 291.





\bibitem{binney}  J. Binney and S. Tremaine, {\it Galactic
Dynamics} (Princeton University Press, Princeton, NJ, 1987).


 \bibitem{verlinde} E. Verlinde,  arXiv:1001.0785 [hep-th];
JHEP 04, 29 (2011).

\end{thebibliography}
\end{document}